\documentclass{iaus}
\usepackage{graphicx}

\title[The Arecibo Galaxy Environemnts Survey]
{The Arecibo Galaxy Environments Survey - Description of the survey and early 
results.}

\author[Minchin et al.]
{R. F. Minchin$^1$, R. Auld$^2$, J. I. Davies$^2$, B. Catinella$^1$,
L. Cortese$^2$, S. Linder$^3$, E. Momjian$^1$, E. Muller$^4$, K. O'Neil$^5$,
J. Rosenberg$^6$, S. Sabatini$^7$, S. E. Schneider$^8$, M. Stage$^8$,
W. van Driel$^9$, the AGES team}

\affiliation{$^1 $Arecibo Observatory, Arecibo, United States,
email: rminchin@naic.edu, 
$^2$ Cardiff University, Cardiff, United Kingdom,
$^3$ University of Hamburg, Hamburg, Germany,
$^4$ Australia Telescope National Facility, Sydney, Australia,
$^5$ National Radio Astronomy Observatory, Green Bank, United States,
$^6$ Harvard Smithsonian Centre for Astrophysics, Cambridge, United States,
$^7$ Osservatorio Astronomico di Roma, Rome, Italy,
$^8$ University of Massachusetts, Amhurst, United States,
$^9$ Observatoire de Paris, Paris, France}

\pubyear{2006}
\volume{235}  
\pagerange{119--119}
\date{17/10/2006}
\jname{Galaxy Evolution across the Hubble Time IAU Symposium 235}
\editors{F. Combes \&  J. Palous, eds.}
\begin{document}

\maketitle

\begin{abstract}
The Arecibo Galaxy Environments Survey (AGES) is a 2000-hour neutral hydrogen 
(H\,{\sc i}) survey using the new Arecibo L-band Feed Array (ALFA) multibeam 
instrument at Arecibo Observatory\footnote{The Arecibo Observatory is part of 
the National Astronomy and Ionosphere Center, which is operated by Cornell 
University under a cooperative agreement with the NSF}. It will cover 200 
square degrees of sky, sampling a range of 
environments from the Local Void through to the Virgo Cluster with higher 
sensitivity, spatial resolution and velocity resolution than previous neutral 
hydrogen surveys.
\end{abstract}


The first field to be covered, 5 square degrees centred on the 
optically-isolated galaxy NGC 1156, has revealed two possible new 
companions to NGC 1156, one of which may be interacting with the galaxy. This 
field also contains 51 definite detections in the volume beyond NGC 1156, 
including one behind a zone of fairly high extinction (1.4 $B$ mag; Schlegel et
al. 1998) to which no optical counterpart
has yet been identified. A further 30 possible sources are currently being
followed up at Arecibo and the GBT.  

\begin{table}[h]
\begin{center}
\caption{Current status of AGES}
\begin{tabular}{lll}\hline
NGC 638 (precursor field)&11/2004 -- 12/2004& Results published in Auld et al. 
(2006)\\
NGC 1156& 12/2005 -- 01/2006& Initial results here, paper in preparation\\
Abell 1367 (only 20\% completed)& 05/2006 -- 06/2006&Initial results in 
Cortese et al. (2007)\\
NGC 7332& 08/2006 -- 09/2006&Currently being analysed\\\hline
\end{tabular}
\end{center}
\end{table}

The status of the survey is given in Table 1 and the full fields are shown
in Fig. 1 (online version only).  An analysis of the noise in the NGC 1156
field is shown in Fig. 2 (online version only) and an analysis of the beam
shape using continuum sources in the cube is shown in Fig. 3 (online version
only).  We find that the noise is Gaussian, with $\sigma = 1$mJy, and that
the final, gridded beam is circular with sidelobes at the 5-10\% level.


\appendix
\section{Online material}

AGES will address a number of scientific objectives including:
\begin{enumerate}
\item the H\,{\sc i} mass function in different environments;
\item baryonic mass density;
\item High Velocity Clouds, dwarf galaxies etc;
\item tidal features;
\item dynamical masses;
\item low column density H\,{\sc i};
\item isolated H\,{\sc i} clouds \& dark galaxies;
\item H\,{\sc i} and QSO absorption features;
\item the spatial distribution of H\,{\sc i} selected galaxies.
\end{enumerate}
In order to meet the scientific goals, AGES will sample the H\,{\sc i}
environment in 13 different fields spread across the sky and sampling
a wide variety of environments (see Fig 1).

\begin{figure}[h]
\resizebox{\columnwidth}{!}{\includegraphics{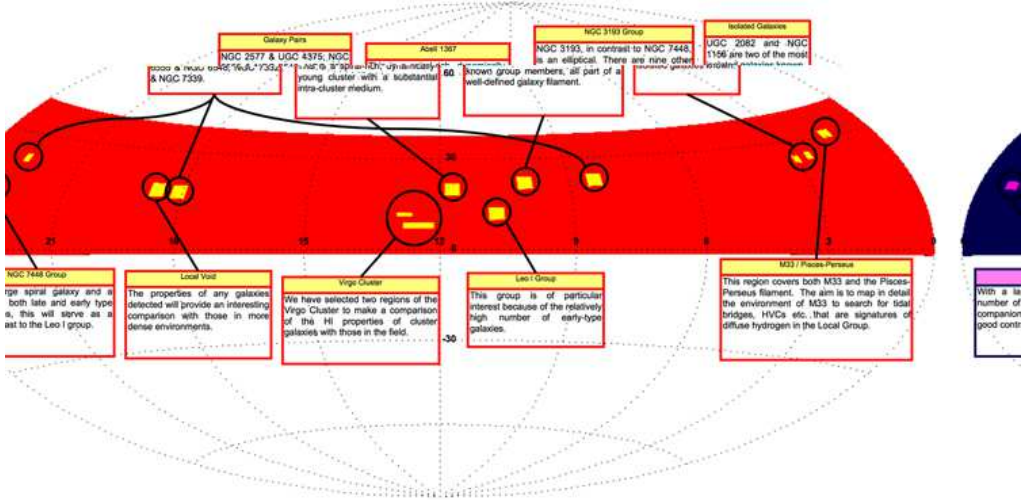}}
\caption{The AGES fields (yellow) and the Arecibo Sky (red).}
\end{figure}

We have analysed the noise levels in the NGC 1156 cube in two ways.  Fig. 2a 
shows how the noise (measured in each RA-Dec plane) varies with velocity.  
There is a
baseline level of around 1 mJy (except at the edges of the cube) with spikes
in the noise at certain velocities.  All of these spikes can be associated
with either radio frequency interference (RFI) or with real sources (see
Table 2).

\begin{table}[h]
\caption{Identification of noise peaks in the NGC 1156 cube}
\begin{tabular}{ll}\hline
H\,{\sc i} in the Milky Way& 0 km\,s$^{-1}$\\
NGC 1156&400 km\,s$^{-1}$\\
AGES J0302+2449&3250 km\,s$^{-1}$\\
Federal Aviation Authority Radar (RFI)& 5800, 7000, \& 15000 km\,s$^{-1}$\\
Global Positioning System Satellites (RFI)&8500 km\,s$^{-1}$\\
WBL 091 Group&10500 km\,s$^{-1}$\\\hline
\end{tabular}
\end{table}

A region free of noise spikes (identified in red in Fig. 2a) was then further
analysed by looking at the distribution of the pixel values.  If the noise
is Gaussian, then this should be a Gaussian distribution.  We find (Fig. 2b)
that the only deviation from the best-fit Gaussian distribution with $\sigma =
0.98$ mJy comes on the positive side of the distribution at around $5\sigma$.
This is consistent with the deviation being due to real sources rather than
a characteristic of the noise.  We therefore conclude that the noise in the
cube is Gaussian in character with a value of around 1 mJy.

\begin{figure}
\resizebox{0.5\columnwidth}{!}{\includegraphics{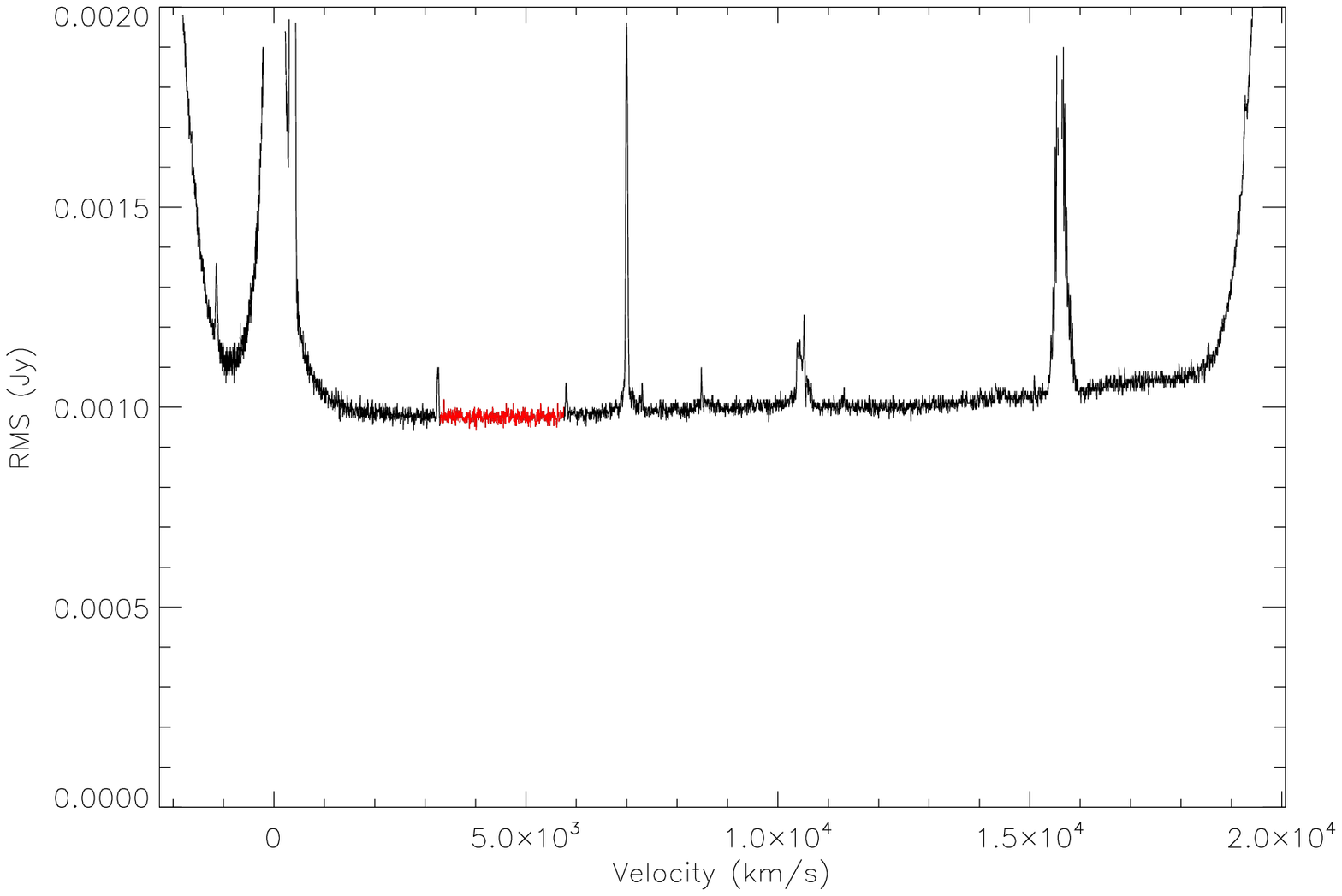}}
\resizebox{0.5\columnwidth}{!}{\includegraphics{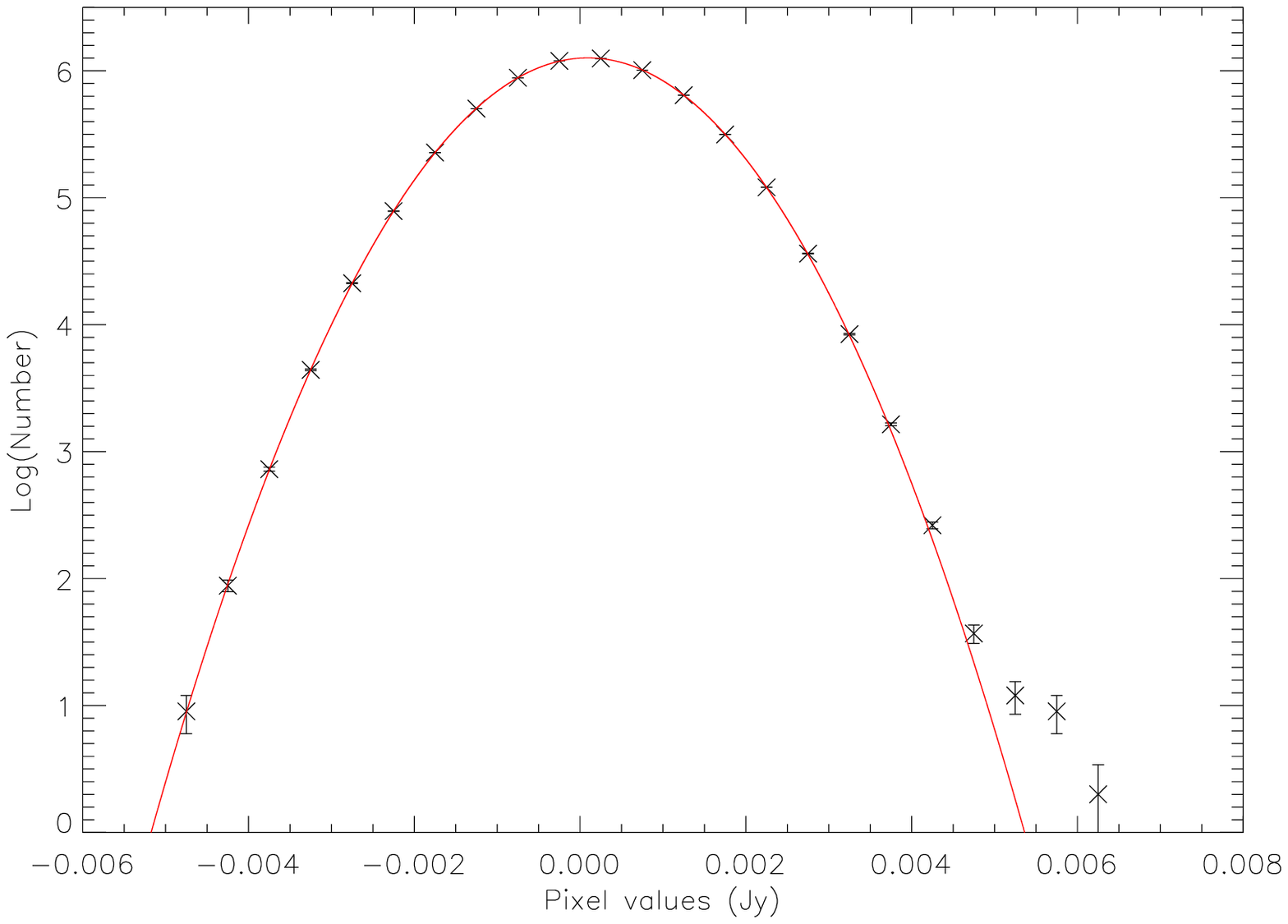}}
\caption{Left: noise in each plane along the cube.  The red
section marks the region used in the noise distribution analysis.  
Right: analysis of the noise distribution in an interference-free section
of the cube; the red line shows a Gaussian distribution with $\sigma = 0.98$ 
mJy.}
\end{figure}

\begin{figure}
\resizebox{0.5\columnwidth}{!}{\includegraphics{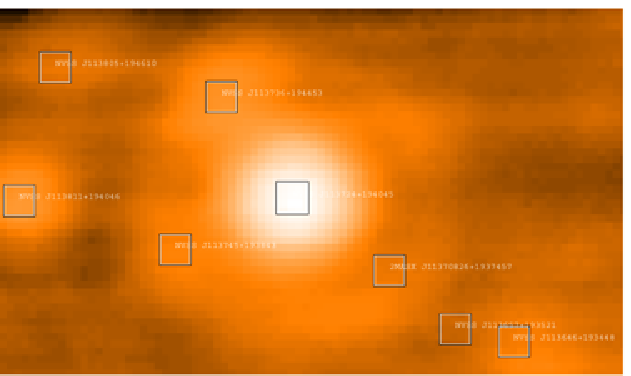}}
\resizebox{0.5\columnwidth}{!}{\includegraphics{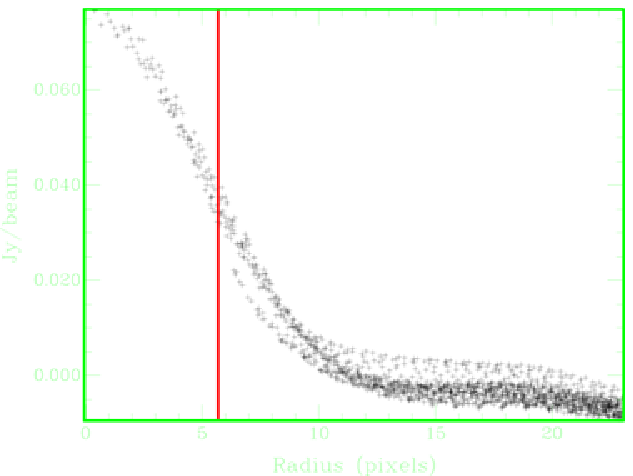}}
\caption{Left: AGES continuum map of the region around NVSS J113724+194045
in the Abell 1367 field, gridded with 0.3 arcmin pixels rather than the 
normal 1 arcmin pixels.  This shows that the beam is fairly close to circular 
with a circular sidelobe pattern.  Other continuum sources detected are marked 
with boxes and labelled.  Right: radial flux distribution around NVSS 
J113724+194045, again with 0.3 arcmin pixels.  The red vertical bar shows 
the size of the theoretical beam
(FWHM 205 arcsec); it can be seen that this corresponds fairly well with
the gridded beam.  The sidelobes fall near the expected position (1.5 $\times$ 
FWHM from the centre) and have a level of 5-10\%.}
\end{figure}

We have also examined the AGES beam shape.  The Arecibo beam is normally 
elliptical and the off-axis ALFA beams also suffer from coma.  The AGES
observing strategy is designed to deal with this by making a Nyquist-sampled
map with every ALFA beam individually.  Thus, when the observations are
median combined, the variations across the beams should be removed, leaving
a circular beam with symmetrical sidelobes.
It can be seen from Fig. 3 that this is indeed the case.  The beam appears
circular with symmetrical sidelobes (Fig. 3a) and the sidelobes are at a
reasonable level (5-10\%) and in the expected position.  The beamsize
is as expected -- the gridding is handled in such a way that the final, gridded
beam is little larger than the raw beam of the telescope.

\end{document}